\documentclass[a4paper]{jpconf}
\usepackage{graphicx}
\usepackage{icsghead}
\begin{document}
\title{Belief propagation with multipoint correlations and its application to Inverse problem}

\author{Masayuki Ohzeki}

\address{Department of Systems Science, Graduate School of Informatics, Kyoto University, }

\ead{mohzeki@i.kyoto-u.ac.jp}

\begin{abstract}
We give explicit formulas of the Bethe approximation with multipoint correlations for systems with magnetic field.
The obtained formulas include the closed form of the magnetization and the correlations between adjacent degrees of freedom.
On the basis of our results, we propose a new iterative algorithm of the improved Bethe approximation.
We confirm that the proposed technique is available for the random spin systems and indeed gives more accurate locations of the critical points.
We discuss the possibility of the application of our method to the Inverse Ising model by use of these equations.
\end{abstract}

\section{Introduction}
Inverse Ising problem, which is originally known as Boltzmann machine learning, has received a lot of attention and been in the active research field.
The problem is to infer both pairwise interactions and external fields of an Ising model given the vector of magnetizations and the matrix of pairwise correlations.
In statistical mechanics, we often deal with the so called forward (direct) process, which consists of the computation of the free energy in order to evaluate several quantities as the magnetization and correlations.
On the other hand, the inverse Ising problem is the backward process.
This can be viewed as the dual counterpart with respect to the direct problem.
The standard approach of the backward process consists of the iterative manipulation to solve the forward process while matching the obtained provisional quantities with the known data.
Therefore we demand relatively fast techniques to compute various quantities in the forward process.

One of the qualitatively poor but very fast approach is the mean-field approximations, which approximately describes a physical system in terms of few parameters. 
A variant of the mean-field approximation, the Bethe approximation, which is originally derived for the ferromagnetic model on regular lattices \cite{Bethe1935,Baxter1982}, has been extended, under the name of cavity method, to models with arbitrary couplings and topologies \cite{Mezard2001}.
Although the Bethe approximation is exact only for tree-like topologies without loops, its application to models defined on random graphs has been very successful asymptotically in large number of components \cite{Mezard2009}.
In order to mitigate the computational cost in the forward process, both of the approximations are often employed.
The mean-field approximation in the inverse Ising problem \cite{Peterson1987,Hinton1989} is very fast but poor description without correlations between the adjacent components (i.e. Ising spins).
By applying the linear response theorem, Kappen and Rodr\`iguez improve the mean-field approximation to give the two-point correlation \cite{Kappen1998}, and its generalization \cite{Tanaka1998} was performed in consistent with the Thouless-Anderson-Palmer approach \cite{Thouless1977} by use of Plefka expansion \cite{Plefka1982}.
In the Bethe approximation, the two-point correlation as well as the expectation of the single component,  which are matched with the given vectors, can be computed in conjunction with the linear response theorem \cite{Tanaka2003,Welling2003,Welling2004,Mezard2009a}.
The Bethe approximation is usually implemented as some iterative algorithm known as the belief propagation \cite{Pearl1988}, which suffers from lack of convergence in complicated system such as the spin glass model.
Therefore it is also important to improve the precision of the mean-field approximation with skillful techniques \cite{Federico2012,Yasuda2013}. 

In recent years, many researchers have attempted to improve the Bethe approximation with correction of the correlation between adjacent degrees of freedom.
One of the approaches is the loop calculus, which is the reduction of the partition function into a summation of contributions from subgraphs that are free of dangling edges \cite{Chertkov2006,Chertkov2006a}. 
Recently it is generalized to the case with multiple solutions of the Bethe approximation \cite{Xiao2011}.
In a different way, a method with correction of loops to the Bethe approximation is proposed by systematically considering correlations between adjacent degrees of freedom \cite{Montanari2005,Parisi2006}.
This technique is a simple to implement an iterative technique as in Ref. \cite{Rizzo2007}.
In the present study, we improve the method to the level with multipoint correlation.
The multipoint correlation was dealt with in the previous study \cite{Rizzo2007} by some approximative procedure, but the explicit form of the correction coming from them was absent.
We give several formulas to be available to compute the various quantities with correction of multipoint correlations.
This is an important step to investigate a nontrivial aspect of the discrepancies between the Bethe approximation and exact calculations of the thermodynamic quantities such as the partition function and free energy.
As an application of our method to the Inverse Ising model, we also give closed equations to estimate the magnetization and pairwise correlations while taking into account the correction of the multipoint correlations.
In addition, we propose an iterative algorithm such as belief propagation to compute the magnetization and correlation on the basis of the obtained formulas.

\section{Standard description of Bethe approximation}
Before going to the central part in this study, it is convenient for readers to find a short review on the Bethe approximation.
Here let us take a very generic instance and explain the motivation to use the Bethe approximation.
We deal with the random-field and random-bond Ising model with the following Hamiltonian
\begin{equation}
H = - \sum_{\langle ij \rangle} J_{ij}S_i S_j - \sum_{i} h_i S_i,
\end{equation}
where $S_i$ is the Ising spin taking $\pm 1$.
The summation denoted by $\langle ij \rangle$ is taken over all pairs of the adjacent spins.
The interaction is denoted as $J_{ij}$, and the magnetic field is represented as $h_i$.
In order to directly calculate the expectation of the local spin and the correlation between adjacent spins, we must obtain the following marginals
\begin{eqnarray}
P(S_i,S_j) &=& \sum_{\{S_k\}/S_i,S_j}P(S_1,S_2,\cdots,S_N)\\
P(S_i) &=& \sum_{S_j}P(S_i,S_j),
\end{eqnarray}
where $P(S_1,S_2,\cdots,S_N)$ is the joint probability of the spin configuration, and $N$ is the number of spins.
The manipulation of the direct evaluation of the marginals (forward process) is intractable in general.
Therefore we often need some approximation to mitigate its difficulty. 
The Bethe approximation is a very reasonable approximation in a moderate computational cost and usually employed to evaluate the marginals in the step of the forward process of the Inverse Ising problem.

There is a qualitatively poorer method than the Bethe approximation, the mean-field approximation.
In the mean-field approximation, the joint probability is regarded as the product of the local distribution of the single spin as
\begin{equation}
P(S_1,S_2,\cdots,S_N) \approx \prod_{i=1}^{N} P_i(S_i).
\end{equation}
The Bethe approximation is a simple improvement of the mean-field approximation by considering the distribution of the two adjacent spins as well as that of the single spins as
\begin{equation}
P(S_1,S_2,\cdots,S_N) \approx \prod_{\langle ij \rangle} \frac{P_{ij}(S_i,S_j)}{P_i(S_i)P_j(S_j)}\prod_{i=1}^{N} P_i(S_i).
\end{equation}
The marginals of the single spin in the denominator exist in order to avoid the multiple count of its effect.

\section{Systematic improvement of Bethe approximation}
\begin{figure}[tb]
\begin{center}
\includegraphics[width=85mm]{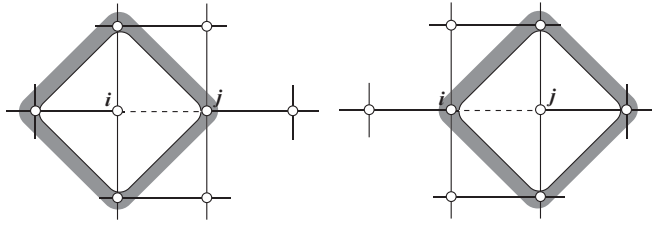}
\end{center}
\caption{{\protect\small sets of the spins adjacent to a particular site $i$ and $j$.
The left (right) panel shows the region that $\partial i$ ($\partial j$) cover with.
We consider the two type of marginals $P_{\partial i}(S_{\partial i})$ and $P_{\partial j}(S_{\partial j})$ in the absence of the interaction between two adjacent spins denoted by $i$ and $j$.
}}
\label{fig1}
\end{figure}

In order to further improve the Bethe approximation, we show a different derivation of the Bethe approximation.
Let us introduce a marginal probability $P_{i}(S_{\partial i})$ for a set of the spins adjacent to a specific site $i$ as depicted in Fig. \ref{fig1}.
We find the exact expressions of the following marginal probabilities for two adjacent spins as
\begin{equation}
P^{(i)}_j (S_i,S_j) = \frac{1}{Z_i}\sum_{S_{\partial i}/j}P_i(S_{\partial i}) \exp\left(\beta H_i S_i + \beta \sum_{l \in \partial i/j}J_{il}S_iS_l\right),
\end{equation}
and 
\begin{equation}
P_i^{(j)}(S_i,S_j) = \frac{1}{Z_j}\sum_{S_{\partial j}/i}P_j(S_{\partial j}) \exp\left(\beta H_j S_j + \beta \sum_{k \in \partial j/i}J_{jk}S_jS_k\right).
\end{equation}
The difference between two marginals stems from the choice of $P_i(S_{\partial i})$ and $P_j(S_{\partial j})$.
We consider the expectation of the local spin by use of the above marginals as
\begin{eqnarray}
A^{(i)}_j &=& \sum_{S_i,S_j} S_i P^{(i)}_j(S_i,S_j)\\
B^{(i)}_j &=& \sum_{S_i,S_j} S_j P^{(i)}_j(S_i,S_j).
\end{eqnarray}
We call the former quantity as the exact cavity magnetization and the latter one as the exact cavity field.
We can relate two of the expectations by the following equation
\begin{eqnarray}
B^{(i)}_j &=& A^{(j)}_i. \label{EBA1}
\end{eqnarray}
This relation holds exactly since any approximations are not employed yet.
Once we obtain the explicit form of two expectations $B^{(i)}_j$ and $A^{(j)}_i$, we can calculate the desired expectations as the magnetizations and correlations.

To connect the above exact cavity magnetization and field to the physical quantities, we consider the partition function.
It is convenient to use the partial partition function instead of the marginal $P_i(S_{\partial i})$ as\begin{equation}
Z_i(S_{\partial i}) = \sum_{\{S_k\}/S_{\partial i},S_i} \exp(\beta \sum_{k\neq i} h_kS_k + \beta \sum_{\langle jk \rangle/\langle\partial i,i\rangle}J_{jk}S_jS_k).
\end{equation}
This is related to the marginal probability $P_i(S_{\partial i})$ as 
\begin{equation}
P_i(S_{\partial i}) = \frac{Z_i(S_{\partial i})}{\sum_{S_{\partial i}} Z_i(S_{\partial i})}.
\end{equation}
Then the partition function can be written as
\begin{equation}
Z = \sum_{S_{\partial i},S_i}\exp(\beta h_i S_i + \beta \sum_{j \in \partial i}J_{ij}S_iS_j) Z_i(S_{\partial i}) .
\end{equation}
In addition, the joint probability of the nearest-neighboring spins is rewritten as
\begin{equation}
P(S_i,S_j) = \frac{1}{Z} \sum_{S_{\partial i/j}}\exp(\beta h_i S_i + \beta \sum_{j \in \partial i}J_{ij}S_iS_j) Z_i(S_{\partial i}) = \frac{\exp(\beta J_{ij}S_iS_j) P_j^{(i)}(S_i,S_j)}{\sum_{S_i,S_j} \exp(\beta J_{ij}S_iS_j) P_j^{(i)}(S_i,S_j)}.
\end{equation}
Notice that these relations hold exactly.

In order to connect the obtained quantity as the cavity field in the Bethe approximation to the desired one, we evaluate the magnetization and disconnected correlation by use of the above exact expressions of the marginals.
It is straightforward to obtain the following expression of the magnetization if one employs the same procedure as the above manipulation in the previous subsection.
\begin{equation}
m_i = \frac{A_j^{(i)} + t_{ij} B_j^{(i)}}{1+ t_{ij} D_{ij}^{(i)}},\label{m_exact}
\end{equation}
where
\begin{equation}
D^{(i)}_{ij} = \sum_{S_i,S_j} S_iS_j P^{(i)}_j(S_i,S_j).
\end{equation}
In addition, the disconnected correlation can be evaluated as
\begin{equation}
\tilde{c}_{ij} = \frac{D^{(i)}_{ij} + t_{ij}}{1+ t_{ij} D_{ij}^{(i)}}.\label{c_exact}
\end{equation}
Equations (\ref{m_exact}) and (\ref{c_exact}) are the exact formulas since any approximations have not yet performed.

In the Inverse Ising problem, the performance of the applied method relies on the precision of the estimation of  the magnetization and correlations.
Therefore we have to improve the precision beyond that of the Bethe approximation.
According to the idea proposed by Montanari and Rizzo \cite{Montanari2005}, let us take the effect of multipoint correlation between adjacent spins to the Bethe approximation.

\subsection{Bethe approximation}
In order to evaluate $B^{(i)}_j$ and $A^{(j)}_i$, we employ some approximation.
One of the simplest way corresponds to the Bethe approximation.

The direct manipulations of three types of the expectations $A^{(i)}_j$, $B^{(i)}_j$, and $D^{(i)}_{ij}$ are 
\begin{eqnarray}
A^{(i)}_j &=& \frac{ t_i \sum_{\mathcal{A}: {\rm even}} t^{(i)}_{\mathcal{A}}\tilde{C}_{\mathcal{A}} + \sum_{\mathcal{A}: {\rm odd}} t^{(i)}_{\mathcal{A}} \tilde{C}_{\mathcal{A}}}{ \sum_{\mathcal{A}: {\rm even}} t^{(i)}_{\mathcal{A}}\tilde{C}_{\mathcal{A} } + t_i \sum_{\mathcal{A}: {\rm odd}} t^{(i)}_{\mathcal{A}} \tilde{C}_{\mathcal{A} }} \\
B^{(i)}_j &=& \frac{ \sum_{\mathcal{A}: {\rm even}} t^{(i)}_{\mathcal{A}}\tilde{C}_{\mathcal{A} \cup j} + t_i \sum_{\mathcal{A}: {\rm odd}} t^{(i)}_{\mathcal{A}} \tilde{C}_{\mathcal{A} \cup j}}{ \sum_{\mathcal{A}: {\rm even}} t^{(i)}_{\mathcal{A}}\tilde{C}_{\mathcal{A} } + t_i \sum_{\mathcal{A}: {\rm odd}} t^{(i)}_{\mathcal{A}} \tilde{C}_{\mathcal{A} }} \\
D^{(i)}_{ij} &=& \frac{ \sum_{\mathcal{A}: {\rm odd}} t^{(i)}_{\mathcal{A}}\tilde{C}_{\mathcal{A} \cup j} + t_i \sum_{\mathcal{A}: {\rm even}} t^{(i)}_{\mathcal{A}} \tilde{C}_{\mathcal{A} \cup j}}{ \sum_{\mathcal{A}: {\rm even}} t^{(i)}_{\mathcal{A}}\tilde{C}_{\mathcal{A} } + t_i \sum_{\mathcal{A}: {\rm odd}} t^{(i)}_{\mathcal{A}} \tilde{C}_{\mathcal{A} }},
\end{eqnarray}
where $\mathcal{A}=\partial i/j$.
Here we define the (disconnected) correlation function for a subset $\mathcal{A}$ as
\begin{equation}
\tilde{C}_{\mathcal{A}}^{(i)} \equiv \sum_{S_{\partial i}}P_i (S_{\partial i}) \prod_{j \in \mathcal{A}} S_j.
\end{equation}
Notice that the connected correlation $C_{\mathcal{A}}^{(i)}$ is related to the above disconnected one as
\begin{equation}
\tilde{C}_{\mathcal{A}}^{(i)} = \sum_{[\mathcal{A}_1,\cdots,\mathcal{A}_n]}C_{\mathcal{A}_1}^{(i)}\cdots C_{\mathcal{A}_n}^{(i)}.
\end{equation}
Here the summation is taken over all possible combinations of the subset $\mathcal{A}$.
We use abbreviations as $t_i = \tanh (\beta h_i)$ and $t^{(i)}_{\mathcal{A}} = \prod_{l \in \mathcal{A}} t_{il}$, where $t_{il} = \tanh(\beta J_{il})$.
In a strict sense, the above correlation should be called the cavity correlation (i. e. the correlation without the specified spin $S_i$).
In the present study, we distinguish the cavity correlation from the standard correlation for clarity.

In the Bethe approximation, we restrict ourselves to the case that two adjacent spins are correlated.
In other words, the adjacent spins to $S_i$ are independent.
Thus $P_i(S_{\partial i}) \approx \prod_{j \in \partial i}P^{(i)}_j({S_j})$.
In this case, the disconnected correlations $\tilde{C}_{\mathcal{A}}^{(i)}$ can be factorized as $\prod_{j \in \mathcal{A}} M_j^{(i)}$.
Here we define the cavity field, which is the expectation of $S_j$ in the absence of the neighboring spin $S_i$, as
\begin{equation}
M_j^{(i)} = \sum_{S_j}S_jP^{(i)}_j({S_j}).
\end{equation}
Then we find the approximate expressions of the expectations as 
\begin{eqnarray}
A^{(i)}_j &=& \frac{ t_i \sum_{\mathcal{A}: {\rm even}}\prod_{j \in \mathcal{A}} t_{ij}M_j^{(i)} + \sum_{\mathcal{A}: {\rm odd}} \prod_{j \in \mathcal{A}} t_{ij}M_j^{(i)} }{ \sum_{\mathcal{A}: {\rm even}} \prod_{j \in \mathcal{A}} t_{ij}M_j^{(i)} + t_i \sum_{\mathcal{A}: {\rm odd}} \prod_{j \in \mathcal{A}} t_{ij}M_j^{(i)} }\\
B^{(i)}_j &=& M_j^{(i)} \\
D^{(i)}_{ij} &=& M_j^{(i)}M_i^{(j)}.
\end{eqnarray}
Therefore we find the self-consistent equation of the cavity field in the Bethe approximation from Eq. (\ref{EBA1}) as
\begin{equation}
M_j^{(i)} = T_j^{(i)}(h_i).\label{BP}
\end{equation}
For simplicity, we define the following quantity
\begin{equation}
T_{j_1,j_2,\cdots,j_k}^{(i)}(h_i) = \tanh\left\{\beta h_i +  \sum_{l \in \partial i/j_1,j_2,\cdots,j_k} \tanh^{-1}\left( t_{il}M_l^{(i)} \right) \right\}.
\end{equation}
In the following, we omit the expression of dependence on the magnetic field unless it appears.
Consequently we obtain the well-known results for the magnetization and correlation at the level of the Bethe approximation as
\begin{eqnarray}
m_i &=& T^{(i)}(h_i) \\
\tilde{c}_{ij} &=& \tanh \left(\beta J_{ij} + \tanh^{-1}(M_j^{(i)}M_i^{(j)})\right).
\end{eqnarray}

\subsection{Improvement of Bethe approximation}
Let us consider the improvement of the Bethe approximation by dealing with the correlation between adjacent spins.
The disconnected cavity correlation is reduced to the product of the cavity field in the Bethe approximation.
We take the effect of the multipoint correlation by considering the disconnected cavity correlation.
For instance, the $3$-body disconnected cavity correlation can be written as
\begin{equation}
\tilde{C}^{(i)}_{jkl} = M^{(i)}_jM^{(i)}_kM^{(i)}_l + C^{(i)}_{jk}M_l^{(i)} + C^{(i)}_{kl}M_j^{(i)} + C^{(i)}_{lj}M_k^{(i)} + C^{(i)}_{jkl}.
\end{equation}
These multipoint cavity correlation can be estimated by the linear response theorem as
\begin{equation}
C^{(i)}_{j,j_1,\cdots,j_k} = \frac{1}{\beta^k}\frac{\partial^k M_j^{(i)}}{\partial h_{j_1}\cdots\partial h_{j_k}}\label{CavCGen}.
\end{equation}
Indeed, the two-point cavity correlation follows
\begin{equation}
C_{j,k}^{(i)} = \left\{1- (M_j^{(i)} )^2 \right\} \left(\delta_{j,k}  + \sum_{k \in \partial j/i} \frac{t_{ik} C^{(j)}_{ik} }{1-t_{ik}^2(M_i^{(j)})^2} \right).\label{CavC}
\end{equation}
This is a kind of the self-consistent equation over the two-point cavity correlations.
Notice that the self-consistent equation consists only of the cavity field and two-point cavity correlations.
The three-point cavity correlations follows a similar equation only with the cavity fields, two-point cavity correlations and themselves.
Once we compute $k$-point cavity correlations, we also estimate the $k+1$-point cavity correlations by successive use of the linear response theorem.

By use of the multipoint cavity correlations, let us consider the correction of the Bethe approximation.
We assume that the multipoint cavity correlation should be small in the following calculation.
We expand several quantities up to the first order of the multipoint correlation in order to evaluate more precise values of the magnetic field and correlation.
The detailed calculation is summarized in Appendix.
The resultant self-consistent equation for the cavity field is 
\begin{eqnarray}\nonumber
M_j^{(i)} &=& T_i^{(j)}(h_j) + \sum_{k=2}\sum_{l_k \in \partial j/i} C_{l_1,\cdots,l_k}^{(j)}\Gamma_{i,l_1,\cdots,l_k}^{(j)}\frac{1-t_j^2}{(1 + t_j T_i^{(j)})^2}\\
& & \quad -  \sum_{k=1}\sum_{j_k \in \partial i/j} C_{j,j_1,\cdots,j_k}^{(i)}\Omega^{(i)}_{j,j_1,\cdots,j_k} ,\label{Update}
\end{eqnarray}
where $\Gamma_{i,l_1,\cdots,l_k}^{(j)}$ and $\Omega^{(i)}_{j,j_1,\cdots,j_k}$ are polynomials consisting of the cavity fields.
This gives a solution $B_j^{(i)}$ (i.e. $A_j^{(i)}$) to the exact relation (\ref{EBA1}) up to the first order of the multipoint cavity correlation function.
In Reference \cite{Montanari2005}, the solution for the case of the homogenous Ising model ($J_{ij}=1$)  without magnetic field up to the first-order of the two-point cavity correlation.
The cavity correlation then was exactly calculated by the Fourier transformation since the system holds translational invariance.
For the two-dimensional case, the infrared divergence avoids the straightforward manipulation to give a explicit solution.

Instead of the analytical approach, in the present study, we propose an iterative algorithm, such as the belief propagation, to find the solution of the improved Bethe approximation.
The algorithmic procedure is as follows.
First, we compute the cavity field by the standard belief propagation.
Then we regard the self-consistent equation (\ref{BP}) as the update recursion for the cavity field.
Second, we recursively estimate the $k$-point cavity correlations (i. e. susceptibility propagation) through the linear response theorem (\ref{CavCGen}) by use of the obtained cavity fields and the resultant $k-1$-point cavity correlations.
For instance, we use Eq. (\ref{CavC}) for the case on the $k=2$-point cavity correlations.
Notice that we keep the cavity fields as the obtained ones in the belief propagation.
Finally, by use of Eq. (\ref{Update}), we calculate the improved cavity field.
We then set the cavity correlations as the obtained ones in the susceptibility propagation.
The improved cavity fields are available for estimation of the magnetization and correlations as
\begin{eqnarray}\nonumber
& & m_i = T^{(i)}(h_i) + \sum_{k=2}\sum_{l_k \in \partial j/i} C_{l_1,\cdots,l_k}^{(j)}\Gamma_{i,l_1,\cdots,l_k}^{(j)}\frac{\left\{1-t_{ij}^2 (M_j^{(i)})^2\right\}(1-t_j^2)}{(1 + t_j T_i^{(j)})^2( 1+ t_{ij}M_j^{(i)}T_i^{(j)}(h_i))^2}\\ \nonumber
& & - \sum_{k=1}\sum_{j_k \in \partial i/j} C_{j,j_1,\cdots,j_k}^{(i)}\Omega^{(i)}_{j,j_1,\cdots,j_k}\frac{t_{ij}(1-(T_j^{(i)})^2)(1-t_it_{ij}M_j^{(i)})\tanh^{\delta_k}(\beta h_i - \tanh^{-1}t_{ij}M_j^{(i)})}{(1+t_iT_j^{(i)})( 1+ t_{ij}M_j^{(i)}T_i^{(j)}(h_i))^2} \\
\label{Magnetization}\\ \nonumber
& & \tilde{c}^{(i)}_{ij} = \tanh \left(\beta J_{ij} + \tanh^{-1}(M_j^{(i)}T_j^{(i)})\right)\\ \nonumber
& & \quad + \sum_{k=2}\sum_{j_k \in \partial i/j} C_{j_1,\cdots,j_k}^{(i)}\Gamma_{j,j_1,\cdots,j_k}^{(i)}\frac{(1-t_{ij}^2)(1-t_i^2)M_j^{(i)}}{(1 + t_i T_j^{(i)})^2( 1 +t_{ij}M_j^{(i)}T_i^{(j)}(h_i))^2}\\
& & \qquad+ \sum_{k=1}\sum_{j_k \in \partial i/j} C_{j,j_1,\cdots,j_k}^{(i)}\frac{(1-t_{ij}^2)(1-t_i^2)(T_j^{(i)})^{1-2\delta_k} \Omega_{j,j_1,\cdots,j_k}^{(i)}}{(1 + t_i T_j^{(i)})^2( 1 +t_{ij}M_j^{(i)}T_i^{(j)}(h_i))^2},\label{Correlation}
\end{eqnarray}
where $\delta_k = 2$ for even $k$ and $\delta_k = 1$ for odd $k$.
\begin{figure}[tb]
\begin{center}
\includegraphics[width=85mm]{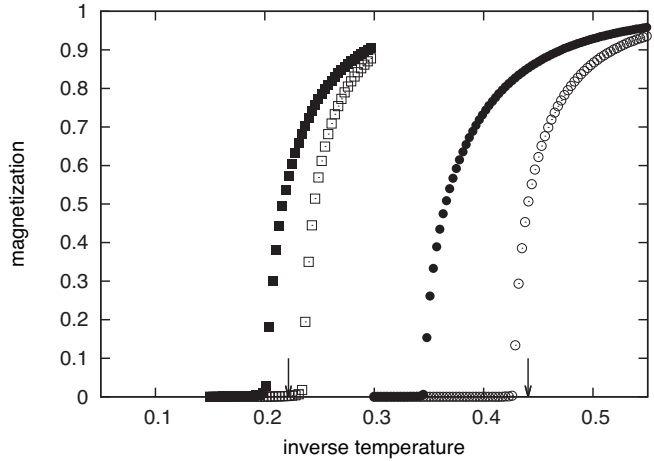}
\end{center}
\caption{{\protect\small Plots of the magnetization for the case of the homogeneous Ising model without magnetic field on the two (white circles) and three-dimensional lattices (white squares).
For comparison, we also describe the results by the ordinary belief propagation (black circles and squares).
The left arrow indicates the existing result ($\beta_c = 0.221 654(6) $\cite{Barber1985}) for the cubic lattice.
The other arrow denotes the exact solution ($\beta_c = 0.440687$ \cite{Kramers1941}) for the square lattice.}}
\label{fig2}
\end{figure}
Equivalently, one may directly insert these results into Eqs. (\ref{m_exact}) and (\ref{c_exact}). 

\begin{figure}[tb]
\begin{center}
\includegraphics[width=85mm]{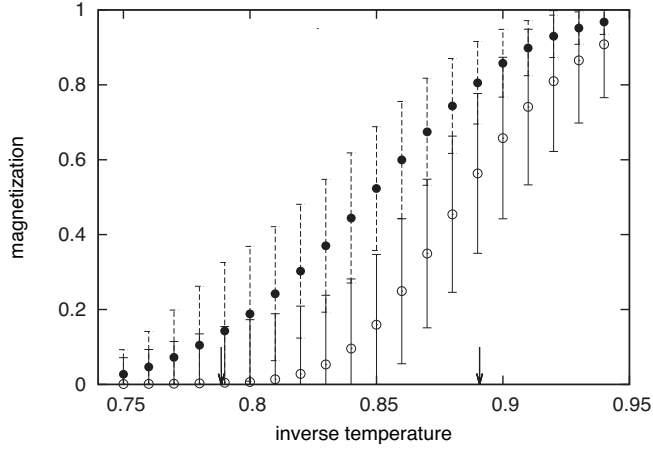}
\end{center}
\caption{{\protect\small Plots of the magnetization for the case of the $\pm J$ Ising model without magnetic field on the two dimensional lattices (white circles).
For comparison, we also describe the results by the ordinary belief propagation (black circles).
The left arrow indicates the estimated result by the Bethe approximation.
The other arrow denotes the expected location given by a precise analysis \cite{Ohzeki2008,Ohzeki2009a}).
We test our algorithm in a small system $L=4$ and use $1000$ samples for $J_{ij}$.}}
\label{fig3}
\end{figure}

We test our algorithm for the homogeneous Ising model without magnetic field  on two and three-dimensional lattice (i.e square and cubic ones).
We first generate $L=24$ lattice and assign the cavity fields and cavity correlations randomly.
Following the our algorithm, we update the cavity fields, and then cavity correlations.
We then obtain the results at the level of the Belief propagation.
Furthermore we compute the corrected estimations following Eq. (\ref{Update}).
In Fig. \ref{fig2}, we show an interesting result of the magnetization for the case of the two-dimensional lattice.
Our iterative method does not suffer from any divergent effect, which avoids the analysis in Ref. \cite{Montanari2005}, to estimate the magnetization for the case of the two-dimensional lattice.
We have not found dependence of the estimations on the linear size of the system.  
The obtained magnetization reveals a closer estimation of the critical point to the exact solution than those of the Bethe approximation.
We also reproduce the result $\beta_c \approx 0.238$ obtained from Eq. (\ref{Magnetization}) in an analytical way \cite{Montanari2005} for the case on the three-dimensional lattice.
The absence of the divergent behavior, which was found in the previous study, stems from the fact that our method deals with both of the effect from the cavity fields and cavity correlations.
The analysis performed in the previous study was under the assumption that the cavity field is very small even in the ferromagnetic phase in the level of the Bethe approximation.
The correction coming from the cavity correlation would compete with effect of increase in the cavity field around the critical point given by the Bethe approximation.
Therefore the sudden change at the wrong critical point as found in the Bethe approximation vanishes, and we observe the critical behavior at another point, which is the corrected critical point.
Our method is available to approximately investigate the spin glass model in the two-dimensional space. 
For the $\pm J$ Ising model on the square lattice with $P(J_{ij}) = p\delta(J_{ij}-1) + (1-p)\delta(J_{ij}+1)$, we test our proposed algorithm to investigate the magnetization across the special critical point, which known as the multicritical point on the Nishimori line \cite{Nishimori1981,Nishimori2001}.
Along the Nishimori line, the inverse temperature is related to the concentration of the ferromagnetic interaction as $\exp(-2\beta) = (1-p)/p$. 
The expected location of the multicritical point is at $p_c \approx 0.8908$ \cite{Ohzeki2008,Ohzeki2009a}.
We find improvement of the critical behavior around a closer point than that given by the belief propagation as shown in Fig. \ref{fig3}
\section{Discussion}
We show the derivation of the generic formulas for the improved Bethe approximation for the case with magnetic field by dealing with correction from not only two-point correlations but also multipoint ones.
Our analysis starts from the exact relationship between cavity fields.
Therefore, if we systematically take the effects of the higher-order multipoint correlations, we asymptotically obtain the exact results of the magnetization and correlations.
Our formulas are available for the Inverse Ising model since we expect precise and fast calculations in the  forward process.
In order to implement our formulas to indeed compute the magnetization and correlations, we propose an iterative algorithm such as the belief propagation.
However this is based on the Bethe approximation to compute the multipoint cavity correlations.
The lack of the convergent behavior in the iterative manipulation, belief propagation, of the Bethe approximation prevents us from investigating a special class of the problems.
Therefore we must develop the method to avoid the poor convergence of the belief propagation to achieve the precise estimations following the current of this study.

In addition, the relationship to the loop calculus should be clarified.
In Ref. \cite{Rizzo2007}, it is pointed out that the direction of the present study and that of the loop calculus might be different.
However the computation by use of the multipoint cavity correlations are also available for estimations of the partition function with loops and reveals the discrepancies from the Bethe approximation.
We hope that the future work shed light on this issue and deepen the understanding of the Bethe approximation and further analysis in statistical mechanics.
\ack
The author thanks the hospitality during the main part of this study in Universit'a di Roma, "La Sapienza".
He also appreciate the fruitful discussions with Tommaso Rizzo, Ulisse Ferrari, Federico Richi-Tersenghi, and Giorgio Parisi.
This work was partially supported by MEXT in Japan, Grant-in-Aid for Young Scientists (B) No.24740263.

\appendix
\section{Correction from cavity correlations}
In order to evaluate the correction of the multipoint correlation, we have to deal with four types of the summation, namely $\sum_{\mathcal{A}: {\rm even}}t_{\mathcal{A}}\tilde{C}^{(i)}_{\mathcal{A}}$, $\sum_{\mathcal{A}: {\rm odd}}t_{\mathcal{A}}\tilde{C}^{(i)}_{\mathcal{A}}$ , $\sum_{\mathcal{A}: {\rm even}}t_{\mathcal{A}}\tilde{C}^{(i)}_{\mathcal{A}\cup j}$ , $\sum_{\mathcal{A}: {\rm odd}}t_{\mathcal{A}}\tilde{C}^{(i)}_{\mathcal{A}\cup j}$.
Notice that $\mathcal{A}$ is a subset except for $S_j$ in the absence of the spin $S_i$.
Let us first write down the explicit form of each quantity up to the first order of the multipoint correlation as $C^{(i)}_{ij}$ and $C^{(i)}_{ijk}$.
\begin{eqnarray}\nonumber
\sum_{\mathcal{A}: {\rm even}}t_{\mathcal{A}}\tilde{C}^{(i)}_{\mathcal{A}} &=& 1 + \sum_{j_1,j_2\in \partial i/j}t_{ij_1}t_{ij_2}\left( M^{(i)}_{j_1}M^{(i)}_{j_2} + C^{(i)}_{j_1,j_2} \right)\\ \nonumber
& & \quad + \sum_{j_k\in \partial i/j}t_{ij_1}t_{ij_2}t_{ij_3}t_{ij_4}\left( M^{(i)}_{j_1}M^{(i)}_{j_2}M^{(i)}_{j_3}M^{(i)}_{j_4} + C^{(i)}_{j_1,j_2}M^{(i)}_{j_3}M^{(i)}_{j_4}  + C^{(i)}_{j_1,j_2,j_3}M^{(i)}_{j_4}\right) \\ \nonumber
& & \qquad + \cdots \\ 
&\equiv& f_0 + \sum_{j_k \in \partial i/j}C_{j_1,j_2}^{(i)}f_{j_1,j_2} + \sum_{j_k \in \partial i/j}C_{j_1,j_2,j_3}^{(i)}f_{j_1,j_2,j_3} + \cdots,
\end{eqnarray}
where $f_0$ is the remaining term in the Bethe approximation, and its explicit form is $\sum_{\mathcal{A}: {\rm even}} t^{(i)}_{\mathcal{A}} \prod_{j \in \mathcal{A}} M_j^{(i)}$.
Here $f_{j_1,\cdots,j_k}$ denotes a function only with the cavity field except for the components appearing in the multipoint cavity correlation $C^{(i)}_{j_1,\cdots,j_k}$.
Similarly the summation of the odd-number components is 
\begin{eqnarray}\nonumber
\sum_{\mathcal{A}: {\rm odd}}t_{\mathcal{A}}\tilde{C}^{(i)}_{\mathcal{A}} &=& \sum_{j_1 \in \partial i/j} t_{ij_1}M_{j_1}^{(i)}\\ \nonumber
& & \quad + \sum_{j_k\in \partial i/j}t_{ij_1}t_{ij_2}t_{ij_3}\left( M^{(i)}_{j_1}M^{(i)}_{j_2}M^{(i)}_{j_3} + C^{(i)}_{j_1,j_2}M^{(i)}_{j_3} + C^{(i)}_{j_1,j_2,j_3}\right) \\ 
&\equiv& g_0 + \sum_{j_k \in \partial i/j}C_{j_1,j_2}^{(i)}g_{j_1,j_2} + \sum_{j_k \in \partial i/j}C_{j_1,j_2,j_3}^{(i)}g_{j_1,j_2,j_3}+ \cdots,
\end{eqnarray}
where $g_0$ is obtained in the Bethe approximation and explicitly written as $\sum_{\mathcal{A}: {\rm odd}} t^{(i)}_{\mathcal{A}} \prod_{j \in \mathcal{A}} M_j^{(i)}$.
We summarize all the terms only with the cavity field except for the components appearing in the multipoint cavity correlation $C^{(i)}_{j_1,\cdots,j_k}$ in a function $g_{j_1,\cdots,j_k}$.
The other two quantities are given in a similar way as
\begin{eqnarray}\nonumber
\sum_{\mathcal{A}: {\rm even}}t_{\mathcal{A}}\tilde{C}^{(i)}_{\mathcal{A}\cup j} &=& M_j^{(i)} f_0 + \sum_{j_k \in \partial i/j} C_{j_1,j_2}^{(i)}f_{j_1,j_2} \\ \nonumber
& & \quad + \sum_{j_k \in \partial i/j}C_{j,j_1}^{(i)}\tilde{f}_{j,j_1} + \sum_{j_k \in \partial i/j} C_{j_1,j_2,j_3}^{(i)}f_{j_1,j_2,j_3} + \sum_{j_k \in \partial i/j}C_{j,j_1,j_2}^{(i)}\tilde{f}_{j,j_1,j_2} + \cdots,\\
\end{eqnarray}
and
\begin{eqnarray}\nonumber
\sum_{\mathcal{A}: {\rm odd}}t_{\mathcal{A}}\tilde{C}^{(i)}_{\mathcal{A}\cup j} &=& M_j^{(i)} g_0 + \sum_{j_k \in \partial i/j}C_{j_1,j_2}^{(i)}g_{j_1,j_2} \\ \nonumber
& & \quad + \sum_{j_k \in \partial i/j}C_{j,j_1}^{(i)}\tilde{g}_{j,j_1} + \sum_{j_k \in \partial i/j} C_{j_1,j_2,j_3}^{(i)}g_{j_1,j_2,j_3}+ \sum_{j_k \in \partial i/j}\sum C_{j,j_1,j_2}^{(i)}\tilde{g}_{j,j_1,j_2},\\
\end{eqnarray}
where we use two functions  $\tilde{f}_{j_1,\cdots,j_k}$ and  $\tilde{g}_{j_1,\cdots,j_k}$ to write the correction of all the terms only with the cavity field except for the components appearing in the multipoint cavity correlation $C^{(i)}_{j_1,\cdots,j_k}$.

For convenience, let us evaluate the ratios of these four quantities.
Then we assume that the multipoint cavity correlations should be small and expand them up to the first order. 
\begin{eqnarray}\nonumber
\frac{\sum_{\mathcal{A}: {\rm odd}}t_{\mathcal{A}}\tilde{C}^{(i)}_{\mathcal{A}}}{\sum_{\mathcal{A}: {\rm even}}t_{\mathcal{A}}\tilde{C}^{(i)}_{\mathcal{A}}} 
&=&
\frac{ g_0 }{ f_0 } + \sum_{j_k \in \partial i/j} C_{j_1,j_2}^{(i)} \frac{ f_0 g_{j_1,j_2} - g_0f_{j_1,j_2}}{ f^2_0 }
\\ & & \quad + \sum_{j_k \in \partial i/j} C^{(i)}_{j_1,j_2,j_3} \frac{ f_0 g_{j_1,j_2,j_3} - g_0f_{j_1,j_2,j_3}}{ f^2_0 }  + \cdots.
\end{eqnarray}
Notice that $f_0$ and $g_0$ appear in the Bethe approximation.
Thus we can obtain 
\begin{eqnarray}
\frac{\sum_{\mathcal{A}: {\rm odd}}t_{\mathcal{A}}\tilde{C}^{(i)}_{\mathcal{A}}}{\sum_{\mathcal{A}: {\rm even}}t_{\mathcal{A}}\tilde{C}^{(i)}_{\mathcal{A}}} &=&
T_j^{(i)} +  \sum_{k=2}\sum_{j_k \in \partial i/j} C_{j_1,\cdots,j_k}^{(i)} \frac{ g_{j_1\cdots,j_k}-  T_j^{(i)} f_{j_1,\cdots,j_k} }{ f_0 } .
\end{eqnarray}
Here we define each coefficient as
\begin{equation}
\Gamma^{(i)}_{j,j_1,\cdots,j_k} \equiv \frac{ g_{j_1,\cdots,j_k}-  T_j^{(i)} f_{j_1,\cdots,j_k} }{ f_0 } .
\end{equation}
Similarly we find
\begin{eqnarray}
\frac{\sum_{\mathcal{A}: {\rm even}}t_{\mathcal{A}}\tilde{C}^{(i)}_{\mathcal{A} \cup j}}{\sum_{\mathcal{A}: {\rm even}}t_{\mathcal{A}}\tilde{C}^{(i)}_{\mathcal{A}}} &=& M_j^{(i)} +  \sum_{k=1}\sum_{j_k \in \partial i/j} C_{j,j_1,\cdots,j_k}^{(i)}\Omega_{j,j_1,\cdots,j_k},
\end{eqnarray}
where
\begin{equation}
\Omega^{(i)}_{j,j_1,\cdots,j_k} \equiv \frac{ \tilde{f}_{j,j_1,\cdots,j_k} }{f_0}.
\end{equation}
In addition,
\begin{eqnarray}
\frac{\sum_{\mathcal{A}: {\rm odd}}t_{\mathcal{A}}\tilde{C}^{(i)}_{\mathcal{A} \cup j}}{\sum_{\mathcal{A}: {\rm even}}t_{\mathcal{A}}\tilde{C}^{(i)}_{\mathcal{A}}} &=& M_j^{(i)}\frac{\sum_{\mathcal{A}: {\rm odd}}t_{\mathcal{A}}\tilde{C}^{(i)}_{\mathcal{A}}}{\sum_{\mathcal{A}: {\rm even}}t_{\mathcal{A}}\tilde{C}^{(i)}_{\mathcal{A}}} +  \sum_{k=1} \sum_{j_k \in \partial i/j}C_{j,j_1,\cdots,j_k}^{(i)} \Upsilon^{(i)}_{j,j_1,\cdots,j_k},
\end{eqnarray}
where we define
\begin{equation}
\Upsilon^{(i)}_{j,j_1,\cdots,j_k} \equiv \frac{\tilde{g}_{j,j_1,\cdots,j_k}}{ f_0 }.
\end{equation}
Each coefficient of the multipoint cavity correlation basically consists of the cavity field.
For instance, let us evaluate $f_{j_1,j_2}/f_0$ in the coefficients
\begin{eqnarray}
\frac{f_{j_1,j_2}}{f_0} &=& t_{ij_1}t_{ij_2}\frac{\sum_{\mathcal{B} = \mathcal{A}/j_1,j_2: {\rm even}}\prod_{j \in \mathcal{B}} t_{ij}M_j^{(i)}}{\sum_{ \mathcal{A}: {\rm even} } \prod_{j \in \mathcal{A}} t_{ij}M_j^{(i)}}.
\end{eqnarray}
Here we use the following identities.
\begin{eqnarray}
E_{j_1,j_2,\cdots,j_k} &=& E_{j_1,j_2,\cdots,j_k,j_l} + t_{il}M_l^{(i)}O_{j_1,j_2,\cdots,j_k,j_l} \\
O_{j_1,j_2,\cdots,j_k} &=& O_{j_1,j_2,\cdots,j_k,j_l} + t_{il}M_l^{(i)}E_{j_1,j_2,\cdots,j_k,j_l},
\end{eqnarray}
where $\sum_{\mathcal{B} = \mathcal{A}/j_1,j_2,\cdots,j_k: {\rm even}}t_{\mathcal{B}}\prod_{j \in \mathcal{B}} M_j^{(i)} = E_{j_1,j_2,\cdots,j_k}$ and $\sum_{\mathcal{B} = \mathcal{A}/j_1,j_2,\cdots,j_k: {\rm odd}}t_{\mathcal{B}}\prod_{j \in \mathcal{B}} M_j^{(i)} = O_{j_1,j_2,\cdots,j_k}$.
Therefore we find 
\begin{eqnarray}
\frac{f_{j_1,j_2}}{f_0} =  t_{ij_1}t_{ij_2}\frac{E_{j_1,j_2}}{E_{j_1,j_2}(1+t_{ij_1}t_{ij_2}M_{j_1}^{(i)}M_{j_2}^{(i)}) + O_{j_1,j_2}(t_{ij_1}M^{(i)}_{j_1}+t_{ij_2}M^{(i)}_{j_2})}.
\end{eqnarray}
Since the ratio of $O_{j_1,j_2,\cdots,j_k}/E_{j_1,j_2,\cdots,j_k}$ is equal to $T^{(i)}_{j,j_1,j_2,\cdots,j_k}$, this equality can be reduced to
\begin{eqnarray}
\frac{f_{j_1,j_2}}{f_0} = \frac{ t_{ij_1}t_{ij_2}}{1+t_{ij_1}t_{ij_2}M_{j_1}^{(i)}M_{j_2}^{(i)} + T^{(i)}_{j,j_1,j_2}(t_{ij_1}M^{(i)}_{j_1}+t_{ij_2}M^{(i)}_{j_2})}.
\end{eqnarray}
In general, we can evaluate $f_{j_1,,\cdots,j_k}/f_0$ from the following recursion. 
\begin{eqnarray}
\frac{f_{j_1,j_2,\cdots,j_{2k},j_{2k+1}}}{f_{j_1,j_2,\cdots,j_{2k}}} &=&  \frac{t_{ij_{2k+1}}T^{(i)}_{j,j_1,\cdots,j_{2k},j_{2k+1}}}{ 1 + t_{ij_{2k+1}}M_{j_{2k+1}}^{(i)}T^{(i)}_{j,j_1,\cdots,j_{2k},j_{2k+1}}} \equiv F_1(j_1,j_2,\cdots,j_{2k},j_{2k+1})\\
\frac{f_{j_1,j_2,\cdots,j_{2k}}}{f_{j_1,j_2,\cdots,j_{2k-1}}} &=& \frac{t_{ij_{2k}}}{ T^{(i)}_{j,j_1,\cdots,j_{2k},j_{2k}} + t_{ij_{2k}}M_{j_{2k}}^{(i)}}\equiv G_1(j_1,j_2,\cdots,j_{2k-1},j_{2k}).
\end{eqnarray}
Similarly we can evaluate the other ratios in the coefficients as
\begin{eqnarray}
& & \frac{\tilde{f}_{j,j_1}}{f_0} = t_{ij_1}\frac{O_{j_1}}{E_{\phi}}= t_{ij_1}\frac{T^{(i)}_{j,j_1}}{1+t_{ij_1}M_{j_1}T^{(i)}_{j,j_1}},
\end{eqnarray}
and
\begin{eqnarray}
\frac{g_{j_1,j_2}}{f_0} &=& t_{ij_1}t_{ij_2}\frac{O_{j_1,j_2}}{E_{\phi}}= t_{ij_1}t_{ij_2}\frac{T^{(i)}_{j,j_1,j_2}}{1+t_{ij_1}t_{ij_2}M_{j_1}^{(i)}M_{j_2}^{(i)} + T^{(i)}_{j,j_1,j_2}(t_{ij_1}M^{(i)}_{j_1}+t_{ij_2}M^{(i)}_{j_2})} \\ 
\frac{\tilde{g}_{j,j_1}}{f_0} &=& t_{ij_1}\frac{E_{j_1}}{E_{\phi}} = t_{ij_1}\frac{1}{1+t_{ij_1}M_{j_1}T^{(i)}_{j,j_1}}.
\end{eqnarray}
For each ratio, we can construct recursion relations similarly to the case of $f$.
In particular, $\tilde{g}_{j,j_1,\cdots,j_{2k+1}}/\tilde{f}_{j,j_1,\cdots,j_{2k+1}} =  1/T^{(i)}_{j,j_1,\cdots,j_{2k+1}}$ and $\tilde{g}_{j,j_1,\cdots,j_{2k}}/\tilde{f}_{j,j_1,\cdots,j_{2k}} = T^{(i)}_{j,j_1,\cdots,j_{2k}}$ holds.
This property ensures 
\begin{eqnarray}
\Omega^{(i)}_{j,j_1,\cdots,j_{2k+1}} &=& T^{(i)}_{j,j_1,\cdots,k_{2k+1}} \Upsilon^{(i)}_{j,j_1,\cdots,j_{2k+1}} \\
\Omega^{(i)}_{j,j_1,\cdots,j_{2k}} &=& \Upsilon^{(i)}_{j,j_1,\cdots,k_{2k}} / T^{(i)}_{j,j_1,\cdots,j_{2k}}.
\end{eqnarray}
As a result, we can find the explicit form of each coefficient.
For instance,
\begin{eqnarray}
\Gamma^{(i)}_{j,j_1,j_2} &=&  \frac{t_{ij_1}t_{ij_2}(T^{(i)}_{j,j_1,j_2} - T^{(i)}_{j})}{1 + t_{ij_1}t_{ij_2}M^{(i)}_{j_1}M^{(i)}_{j_2} + t_{ij_2}M^{(i)}_{j_1}T^{(i)}_{j,j_1,j_2} + t_{ij_1}M^{(i)}_{j_2}T^{(i)}_{j,j_1,j_2}}\\
\Omega^{(i)}_{j,j_1}  &=& \frac{t_{ij_1}T_{j,j_1}^{(i)}}{1+t_{ij_1}M_{j_1}^{(i)}T_{j,j_1}^{(i)}} = T_{j,j_1}^{(i)}\Upsilon^{(i)}_{j,j_1}.
\end{eqnarray}
Once we find these coefficients, we can reproduce the previous results explicitly given by Montanari and Rizzo \cite{Montanari2005}.
By the present analysis, we push up their to the further improvement with multipoint correlation as\begin{eqnarray}
A^{(i)}_j &=& T_j^{(i)}(h_i)+ \sum_{k=2}\sum_{j_k \in \partial i/j} C_{j_1,\cdots,j_k}^{(i)}\Gamma_{j_1,\cdots,j_k}^{(i)}\frac{1-t_i^2}{(1 + t_i T_j^{(i)})^2}\\
B^{(i)}_j &=& M_j^{(i)} + \sum_{k=1}\sum_{j_k \in \partial i/j} C_{j,j_1,\cdots,j_k}^{(i)} \Omega^{(i)}_{j,j_1,\cdots,j_k} .
\end{eqnarray}
We thus obtain Eq. (\ref{Update}).
We reproduce the same result for the case without magnetic field and only with $C_{j_1,j_2}^{(i)}$ and $C^{(i)}_{j,j_1}$ as that given in Ref. \cite{Montanari2005}.
Indeed the update equation can estimate closer locations of the critical point to its exact answer that those by the Bethe approximation.
In order to evaluate two-point correlation to match with the given vector of the data in the case of the Inverse Ising model, we evaluate $D^{(i)}_{ij}$ from the above formulas.
The result is 
\begin{eqnarray}\nonumber
D^{(i)}_{ij} &=& T_j^{(i)}(h_i)M_j^{(i)} + \sum_{k=1}\sum_{j_k \in \partial i/j} C_{j,j_1,\cdots,j_k}^{(i)} \Omega^{(i)}_{j,j_1,\cdots,j_k}\frac{(T_j^{(i)})^{d_k}}{1+t_iT_j^{(i)}} \\
& & \quad +\sum_{k=2}\sum_{j_k \in \partial i/j} C_{j_1,\cdots,j_k}^{(i)}\Gamma_{j_1,\cdots,j_k}^{(i)}\frac{(1-t_i^2)M_j^{(i)}}{(1 + t_i T_j^{(i)})^2}.
\end{eqnarray}
and obtain Eqs. (\ref{Magnetization}) and (\ref{Correlation}).
Equivalently, one may directly insert these results into Eqs. (\ref{m_exact}) and (\ref{c_exact}). 

\section*{References}
\bibliography{paper}

\end{document}